\documentclass[onecolumn,showpacs,preprintnumbers,amsmath,amssymb]{revtex4}

\usepackage{graphicx}
\usepackage{dcolumn}
\usepackage{bm}

\begin{document}


\title{Understanding reversible looping kinetics of a long polymer molecule in solution. Exact solution for two state model with delta function coupling}

\author{Moumita Ganguly}
\email{mouganguly09@gmail.com} 
\author{Aniruddha Chakraborty}
\affiliation{School of Basic Sciences, Indian Institute of Technology Mandi,
Kamand Campus, Himachal Pradesh 175005, India. }

\date{\today}

\begin{abstract}
In this paper, the looping kinetics of a long polymer chain in solution has been investigated by using two state model, where one state represents open chain polymer molecule and the other represents closed chain polymer molecule. The dynamics of end-to end distance of both open chain and closed chain polymer is represented by a Smoluchowski-like equation for a single particle under two different harmonic potentials. The coupling between these two potentials are assumed to be a Dirac Delta function in our model. We evaluate two rate constants, the long term and the average rate constant.  We have also incorporated the effect of all other chemical reactions involving at least one end of the open chain polymer molecule - on rate of end-to-end loop formation. The closed chain polymer molecule can be converted to a open chain molecule by breaking of any bond - the effect of this reaction on the rate of end-to-end loop formation is also considered in our model.   
\end{abstract}

\maketitle
\noindent Reversible looping of a long chain polymer molecule in solution is an interesting problem \cite{Wilemski}. A large number of theoretical and experimental studies have been done already in this area \cite{Cheriyl, Bonnet, Goddard}. In this paper, our model of reversible looping dynamics of a long polymer molecule have been formulated following the work of Szabo {\it et. al.,} \cite{Schulten}. In our model the dynamics of end-to-end distance of an open polymer chain in solution is mathematically represented by a Smoluchowski-like equation for a single particle under harmonic potential and the dynamics of end-to-end distance of a closed polymer chain is represented by the same Smoluchowski equation but with a different harmonic potential. Also, one of the high point of our model is incorporating the effect of rates of all other chemical reactions involving at least one end of the polymer -  on the end-to-end looping rate and another is the fact that the closed chain polymer molecule can be converted to an open chain by breaking any bond - effect of this reaction on the rate of looping is also considered in our model.
\par
\noindent The most simplest one dimensional description of probability distribution $P_{o}(x,t)$ of end-to-end distance of a long open chain polymer at time $t$  is given by \cite{Szabo}, 
\begin{equation}
\frac{\partial P_{o}(x,t)}{\partial t} = \left(\frac{4 N b^2}{\tau_{o}} \frac{\partial^2}{\partial x^2} + \frac{2}{\tau_{o}} \frac{\partial}{\partial x} x \right) P_{o}(x,t),
\end{equation}
where $x$ denotes end-to-end distance. $\tau_{o}$ is the relaxation time from one to another configuration, length of the polymer is given be $N$, `$b$' denotes the bond length of the polymer and $x$ denotes the end-to-end distance. In this model, one can incorporate the effect of all other chemical reactions (involving at
least one of the end group) apart from the end-to-end loop formation adding  $- k_{s} P_{o}(x, t)$ term on the R.H.S. of the above equation \cite{Mou-1}.
\par
\begin{equation}
\frac{\partial P_{o}(x,t)}{\partial t} = \left(\frac{4 N b^2}{\tau_{o}} \frac{\partial^2}{\partial x^2} + \frac{2}{\tau_{o}} \frac{\partial}{\partial x} x - k_{s}\right) P_{o}(x,t),
\end{equation}
\noindent  Now if the two ends of the polymer molecule meet, a loop would be form. Once formed,  the end-to-end distance of that closed chain polymer will also evolve in time according to the following equation
\begin{equation}
\frac{\partial P_{c}(x,t)}{\partial t} = \left(\frac{4 N b^2}{\tau_{c}} \frac{\partial^2}{\partial x^2} + \frac{2}{\tau_{c}} \frac{\partial}{\partial x} x \right) P_{c}(x,t).
\end{equation} 
In the above $P_{c}(x,t)$ denotes the probability distribution of end-to-end distance of a closed long chain polymer at time $t$ and $\tau_{c}$ is the relaxation time from one to another configuration. In the above equation, one can incorporate the effect of all bond breaking chemical reactions (other than the so called 'end-to-end bond' breaking reaction)  adding  $- k_{s} P_{c}(x, t)$ term on the R.H.S. of the above equation
\begin{equation}
\frac{\partial P_{c}(x,t)}{\partial t} = \left(\frac{4 N b^2}{\tau_{c}} \frac{\partial^2}{\partial x^2} + \frac{2}{\tau_{c}} \frac{\partial}{\partial x} x - k_{s}\right) P_{c}(x,t).
\end{equation} 
The interconversion of open and  closed configurations can be can be understood by using the following equations.
\begin{eqnarray}
\frac{\partial P_{o}(x,t)}{\partial t} = \left(\frac{4 N b^2}{\tau_{o}} \frac{\partial^2}{\partial x^2} + \frac{2}{\tau_{o}} \frac{\partial}{\partial x} x - k_{s}\right) P_{o}(x,t) - \frac{S(x)}{\tau_{o}} P_{c}(x,t)\\ \nonumber
\frac{\partial P_{c}(x,t)}{\partial t} = \left(\frac{4 N b^2}{\tau_{c}} \frac{\partial^2}{\partial x^2} + \frac{2}{\tau_{c}} \frac{\partial}{\partial x} x - k_{s}\right) P_{c}(x,t) - \frac{S(x)}{\tau_{o}} P_{o}(x,t). \nonumber
\end{eqnarray}  
In the above equation $S(x)$ represents position dependent coupling term (taken to be normalized {\it i.e.} $\int_{-\infty}^{\infty} S(x)dx = 1$, couples open chain and closed chain configuration).
\noindent Now we do the Laplace transform $\tilde P_{i}(x,s)$ of $P_{i}(x,t)$ by \\
\begin{equation}
\tilde P(x,s)= \int^\infty_0 P(x,t) e^{-st} dt.
\end{equation}
Laplace transformation of Eq.(5) gives\\
\begin{eqnarray}
\left[s  - {L_{o}}+k_{s}\right] {\tilde P_{o}}(x,s) + \frac{S(x)}{\tau{o}}{\tilde P_{c}}(x,s)=  P_{o}(x,0)\\ \nonumber
\left[s  - { L_{c}}+k_{s}\right] {\tilde P_{c}}(x,s) + \frac{S(x)}{\tau_{c}}{\tilde P_{o}}(x,s)=  P_{c}(x,0),\nonumber
\end{eqnarray}
where $P_{o}(x,0)$ is the initial end-to-end probability distribution of the open chain polymer and $P_{c}(x,0)=0$ is the initial end-to-end probability distribution of the closed chain polymer. Also in the above equation ${L_{i}}$ is defined as follows
\begin{equation}
{L_{i}} = \frac{4 N b^2}{\tau_{i}} \frac{\partial^2}{\partial x^2} + \frac{2}{\tau_{i}} \frac{\partial}{\partial x} x.
\end{equation}
In the following we assume $P_{c}(x,0)=0$ and we write Eq.(6) in the matrix form 
\begin{equation}
 \left(
\begin{array}{c}
P_{o} (x,s) \\
P_{c} (x,s)
\end{array}
\right) = \left(
\begin{array}{cc}
s - L_{o}+ k_{s} & \frac{S(x)}{\tau_{o}} \\
\frac{S(x)}{\tau_{c}} & s - L_{c}+ k_{s}
\end{array}
\right)^{-1}
\left(
\begin{array}{c}
P_{o}(x) \\
0
\end{array}
\right).
\end{equation}
Using the partition technique \cite{Lowdin}, solution of this equation may be expressed as 
\begin{equation}
{\tilde P}_{o}(x,s)=\int_{-\infty}^{\infty} dx_0 G(x,s;x_0)P_{o}(x_0),
\end{equation}
where $G(x,s;x_0)$ is the Green's function given by
\begin{equation}
G(x,s;x_0)=\left < x \left|[s- L_{o}+ k_{s} + {k_0}^2 \frac{S}{\tau_o}[s- L_c+k_s]^{-1}\frac{S}{\tau_c}]^{-1}\right| x_0 \right>.
\end{equation}
The above equation is true for any general $S(x)$. This expressions simplify a lot, if $S(x)$ is a Dirac Delta function located at $x=0$. Then
\begin{equation}
G(x,s;x_0)=\left < x \left|[s- L_o+ k_{s} + \frac{{k_0}^2}{\tau_c \tau_o} G^{0}_c(0,s;0) S ]^{-1}\right| x_0 \right>,
\end{equation}
where
\begin{equation}
G^{0}_c(x,s;x_0)=\left < x \left|[s-L_c+ k_{s} ]^{-1}\right| x_0 \right>
\end{equation}
and corresponds to the time evolution of the end-to-end distance of a closed chain long polymer starting from $x_0$ in the absence of any coupling.
Now we use the operator identity
\begin{eqnarray}
\left[s- L_o + k_s -\frac{{k_0}^2}{\tau_c \tau_o} G^{0}_c(0,s;0) S\right]^{-1}=\left[s-L_o+ k_s\right]^{-1}- \left[s- L_o+ k_s\right]^{-1} \times \\ \nonumber
\frac{{k_0}^2}{\tau_c \tau_o} G^{0}_c(0,s;0) S \left[s- L_c + k_s - \frac{{k_0}^2}{\tau_c \tau_o} G^{0}_c(0,s;0) S \right]^{-1}.\nonumber
\end{eqnarray}
to get
\begin{eqnarray}
\left< x \left|\left[s- L_o + k_s\textbf{} -\frac{{k_0}^2}{\tau_c \tau_o} G^{0}_c(0,s;0) S\right]^{-1}\right| x_0 \right> = \left< x \left|\left[s-L_o+ k_s\right]^{-1} \right| x_0 \right> \\ \nonumber 
- \left< x \left|\left[s- L_o+ k_s\right]^{-1} \frac{{k_0}^2}{\tau_c \tau_o} G^{0}_c(0,s;0) S \left[s- L_c + k_s - \frac{{k_0}^2}{\tau_c \tau_o} G^{0}_c(0,s;0) S \right]^{-1}
\right| x_0 \right>.\nonumber
\end{eqnarray}
Inserting the resolution of identity $I=\int_{-\infty}^{\infty} dy \left|y \left> \right < y \right|$ in the second term of the above equation, we arrive at the following equation.
\begin{equation}
G(x,s;x_0)=G^0_{o}(x,s;x_0) - \frac{{k_0}^2}{\tau_c \tau_o} G^0_o(x,s;0)G^0_c(0,s;0)G(0,s;x_0),
\end{equation}
where $G^0_{o}(x,s;x_0)=\left < x \left|[s-L_o+k_s]^{-1}\right| x_0 \right>$ corresponds to the time evolution of the end-to-end distance of an open chain long polymer starting from $x_0$ in the absence of any coupling.  We now put $x=0$ in the above equation and solve for $G(0,s;x_0)$ to get
\begin{equation}
G(0,s;x_0)=\frac{G^0_{o}(0,s;x_0)}{1+{k_0}^2 G^0_o(0,s;0)G^0_c(0,s;0)}.
\end{equation}
This when substitued back into Eq. (16) gives
\begin{equation}
G(x,s;x_0)=G^0_o(x,s;x_0) - \frac{\frac{{k_0}^2}{\tau_c \tau_o} G^0_o(x,s;0)G^0_c(0,s;0)G^0_o(0,s;x_0)}{1+\frac{{k_0}^2}{\tau_c \tau_o} G^0_o(0,s;0)G^0_c(0,s;0)}.
\end{equation}
In the above $G(x,s;x_0)$ is expressed in terms of Green's function $G^{0}_{o}(x,s|x_0)$ and $G^{0}_{c}(x,s|x_0)$ and corresponds to change in end-to-end distance of the open or closed chain polymer, that has the inceptive value $x_0$, in the absence of any sink. It is noteworthy that Laplace transform of $g^0_{i}(x,t|x_0)$ gives the probability that the end-to-end distance of a open or closed chain polymer per say, starting at $x_0$ may be found at $x$, at time $t$. It obeys the following equation,
\begin{equation}
\left[(\partial /\partial t)- L_{i} \right] g^0_{i}(x,t|x_0)=\delta(x-x_0),
\end{equation}
where for open chain polymer `i' is nothing but `o' and for closed chain polymer `i' is nothing but `c'. The above equation doesn't have a sink term in it. In the absence of sink, there is no possibility of end-to-end bond formation for open chain polymer or no possibility of end-to-end bond breaking for closed chain polymer . Therefore, $\int^\infty_{-\infty} dx g^0_{i}(x,t|x_0) = 1$. From this we can conclude 
\begin{equation}
\int^\infty_{-\infty} dx G^0_{i}(x,s|x_0) = 1/s
\end{equation}
Using the expression of $G(x,s|x_0)$ in Eq. (9) we get ${\tilde P}_{o}(x,s)$ explicitly. It is difficult to calculate survival probability $P_o(t) =\int^\infty_{-\infty} dx P_o(x,t)$. Instead one can easily calculate the Laplace transform ${\tilde P}_o(s)$ of $ P_o(t)$ directly. ${\tilde P}_o(s)$ is associated to ${\tilde P}_o(x,s)$ by 
\begin{equation}
{\tilde P}_o(s) = \int^\infty_{-\infty} dx {\tilde P}_{o}(x,s).
\end{equation} 
From Eq. (9), Eq. (17) and Eq. (20), we get
\begin{equation}
{\tilde P}_o(s)=\frac{1}{s+k_{s}}\left[1-[1+\frac{k_0^2}{\tau_{c} \tau{o}} G^0_c(0,s;0) G^{0}_o(0,s+k_{s}|0)]^{-1} \frac{k_0^2}{\tau_{c}\tau_{o}} G^0_c(0,s;0)\times \int^\infty_{-\infty} dx_0 \; G^{0}_{o} (0,s+k_{s}|x_0)P(x_0,0).\right]
\end{equation}
The average and long time rate constants can be derived from $ {\tilde P}_o(s).$ Thus, $k^{-1}_I ={\tilde P}_o(0)$ and $k_L$ = negative of the pole of ${\tilde P}_o(s),$ which is close to the origin. From (21), we obtain
\begin{equation}
k^{-1}_I = \frac{1}{k_{s}}\left[1-[1+\frac{k_0^2}{\tau_{c} \tau{o}} G^0_c(0,0;0) G^{0}_o(0,k_{s}|0)]^{-1} \frac{k_0^2}{\tau_{c}\tau_{o}} G^0_c(0,0;0)\times \int^\infty_{-\infty} dx_0 \; G^{0}_{o} (0,k_{s}|x_0)P(x_0,0).\right]
\end{equation}
Thus $k_I$ depends on the initial probability distribution $P(x,0)$ whereas $k_L = - $ pole of $[\;[ 1+\frac{k_0^2}{\tau_{c} \tau{o}}\; G^0_c(0,s;0) G^{0}_{o}(0, s+k_s|0)][s+k_s]\;]^{-1}$, the one which is closest to the origin, on the negative $s$ - axis, and is independent of the initial distribution $P(x_0,0)$.
The $G^{0}_{i}(x,s;x_0)$ can be found out by using the following equation
\begin{equation}
\left(s - L_{i}\right) G^{0}_{i}(x,s;x_0)= \delta (x - x_0) 
\end{equation}
Using standard method \cite{Hilbert} to obtain.
\begin{equation}
G^{0}_{i}(x,s;x_0)=F_{i}(z,s;z_0)/(s+k_s)
\end{equation}
with
\begin{equation}
F_{i}(z,s;z_0)= D_{\nu_i}(-z_<)D_{\nu_i}(z_>)e^{(z_0^2-z^2)/4}\Gamma(1-{\nu_i})[1/(4 \pi N b^2)]^{1/2} 
\end{equation}
In the above, $z$ defined by $z = x(2Nb^2)^{1/2}$  and $z_j = x_j(2Nb^2)^{1/2}$, $\nu_{i} = —s{\tau_{i}}/2$ and $\Gamma(\nu_{i})$ is the gamma function. Also, $z_{<}= min(z, z_0)$ and $z_{>}= max(z, z_0)$. $D_{\nu_{i}}$ represent parabolic cylinder functions. To get an understanding of the behavior of $k_I$ and $k_L$, we assume the initial distribution $P^0_e(x_0)$ is represented by $\delta(x-x_0)$. Then, we get 
\begin{equation}
{k_I}^{-1}= {k_s}^{-1}\left(1 - \frac{\frac{k_0^2}{\tau_{o} \tau{c}} G^0_c(0,0;0) F_{o}(0,k_s|z_0)}{k_s+ {\frac{k_0^2}{\tau_{c}\tau_{o}}}G^0_c(0,0;0) F_{o}(0,k_s|0)} \right)
\end{equation}
Again
\begin{equation}
k_L= k_s - [ values \; of \; s \; for \; which \;\; s+ {\frac{k_0^2}{\tau_{c} \tau{o}}}G^0_c(0,s;0)F_{o}(z_s,s|z_s)=0]
\end{equation}
We should mention that $k_I$ is dependent on the initial position $x_0$ and $k_s$ whereas $k_L$ is independent of the initial position.
In the following, we assume $k_s\rightarrow$ 0, in this limit we arrive at conclusions, which we expect to be valid even when $k_s$ is finite. Using the properties of $D_{v_i}{(z)}$, we find that when $k_s\rightarrow 0, F{(0,k_s|z_0)}$ and $F{(0,k_s|0)}\rightarrow {[1/(4\pi Nb^2)]}^\frac{1}{2}$so that
\begin{equation}
\frac{k_0^2}{\tau_{c} \tau_{o}} G^0_c(0,s;0) F{(0,k_s|z_0)}/[k_s+ \frac{k_0^2}{\tau_c \tau_{o}}G^0_c(0,s;0)F{(0,k_s|0)}]\rightarrow 1.
\end{equation}
\\Hence 
\begin{equation}
k_I^{-1}=-{[\frac{\partial}{\partial k_s}\left[\frac{\frac{k_0^2}{\tau_{c} \tau_{o}} G^0_c(0,s;0) F(0,k_s|z_0)}{k_s + \frac{k_0^2}{\tau_c \tau_o} G^0_c(0,s;0) F(z_s,k_s|z_s)}\right]}_{k_s \rightarrow 0}
\end{equation} 
If we take $z_0 < 0 $, so that the particle is initially placed to the left of sink. Then \\
\begin{equation}
k_I^{-1}= \frac{\tau_o \tau_c}{k_0^2 G^0_c(0,s;0)}{[1/{(4\pi N b^2)}]}^{1/2}+ \left[\frac{\partial}{\partial k_s}\left[\frac{e^{[z_0^2/4]}D_{v_o}{(-z_0)}}{D_{v_o}{(0)}}\right]\right]_{v=0}
\end{equation} 
After simplification
\begin{equation}
k_I^{-1}= \frac{\tau_c \tau_o}{k_0^2 G^0_c(0,s;0)}{[1/{(4\pi N b^2)}]}^{1/2}+ \left(\int_{z_0}^{0} dz e^{(z^2/2)}\left[1+erf(z/\sqrt{2}\right]\right)(\pi/2)({\tau_o/2})
\end {equation}
The long-term rate constant $k_L$ is determined by the value of $s$, which satisfy $s+ \frac{k_0^2 G^0_c(0,s;0)}{\tau_o \tau_c}F(0,s|0)=0$. This equation can be written as an equation for $\nu_o (= -s{\tau_o} /2)$
\begin{equation}
\nu_o = D_{\nu_o}(0) D_{\nu_o}(0)\Gamma(1-\nu_{o})\frac{\frac{k_0^2 G^0_c(0,s;0)}{\tau_{o}\tau_{c}}} {4 b \sqrt{\pi N}} 
\end{equation}
For integer values of $\nu_{i}$, $D_{\nu_i}(z)=2^{-\nu_{i}/2}e^{-z^2/4}H_{\nu_{i}}(z/\sqrt{2})$, $H_{\nu_i}$ are Hermite polynomials. $\Gamma(1-\nu_{i})$ has poles at $\nu_{i} = 1,2, . . . .$. 

Our interest is in $\nu_i \in[0, 1]$, as $k_L = \frac{2}{\tau_o} \nu_o$ for $k_s =0$. If 
$ \frac{k_0^2 G^0_c(0,s;0)}{(\tau_{o} \tau{c} 4 b \sqrt{\pi N})}  \ll 1$, or
$z_c \gg 1$ then $\nu_i \ll 1$ and one can arrive
\begin{equation}
\nu_o = D_0(0)D_0(0)\frac{k_0^2 G^0_c(0,s;0)}{\tau+c \tau_o(4 b \sqrt{\pi N)}} 
\end{equation}
and hence 
\begin{equation}
k_L = \frac{ \tau_c \tau_o}{k_0^2 G^0_c(0,s;0)}{[1/{(4\pi N b^2)}]}^{1/2}
\end{equation}

\noindent In this paper we give a very simple two state model for understanding the kinetics of reversible looping of long polymer chain in solution. 
Explicit expressions for $k_I$ and $k_L$ have been derived. Our model takes care of effect of all other chemical reactions involving at least one of ends of the polymer, on the end-to-end reaction rate. We also incorporate the effect of breaking of any bond of closed chain polymer molecule, on the rate end-to-end loop formations. 

\begin{acknowledgments}
\noindent One of the author (M.G.) would like to thank IIT Mandi for HTRA fellowship and the other author thanks IIT mandi for providing CPDA grant.
\end{acknowledgments}

\end{document}